\documentclass[preprint2]{aastex}

\begin{document}
\title{Electrodynamics of Massless Charges with Application to Pulsars} 
\author{Andrei Gruzinov}
\affil{New York University Abu Dhabi, near Sheikh Khalifa Mosque,  Abu Dhabi, United Arab Emirates }

\begin{abstract}

Electromagnetic field together with zero-mass charges moving in this field form a well-behaved semi-dissipative dynamical system -- Electrodynamics of Massless Charges (EMC). We give equations of EMC,  argue that  EMC is an adequate theory for calculating pulsar magnetospheres, give an illustrative numerical calculation (showing that bolometric luminosity of an aligned rotator is approximately equal to half the spin-down power). EMC looks like a portion of the full pulsar theory that will resolve the already calculated  bolometric luminosity into light curves and spectra.

~

~

~

\end{abstract}

\section{ Electrodynamics of Massless Charges (EMC)}

EMC postulates that positive and negative charges move with unit, meaning speed of light, velocities:
\begin{equation}\label{vel}
{\bf v}_\pm={{\bf E}\times {\bf B}\pm(B_0{\bf B}+E_0{\bf E})\over B^2+E_0^2}.
\end{equation}
Here $E_0$ and $B_0$ are the proper electric field scalar and the proper magnetic field pseudo-scalar, defined by
\begin{equation}\label{e0b0}
B_0^2-E_0^2=B^2-E^2,~~ B_0E_0={\bf B}\cdot {\bf E}, ~~E_0\geq 0.
\end{equation}
Equations (\ref{e0b0}) mean that $E_0$ and $|B_0|$ are the field magnitudes in the frame where the fields are parallel, with the sign of $B_0$ negative when the fields are antiparallel.

The "Aristotelean" dynamics (\ref{vel}), with the particle velocity rather than acceleration given by the force, results from an infinite radiation damping of massless charges.  Equation (\ref{vel}) means that in the frame where the fields are parallel, charges move along the common direction of the fields at the speed of light. Moving along the fields nullifies the leading-order radiation damping term. And, because of curvature radiation, the parallel dynamics is also instantaneous, fixing the velocities at (\ref{vel})\footnote{The terminal Lorentz factor due to curvature radiation is given by $\gamma ^4\sim E_0R^2/e$, where $R$ is the radius of curvature of the trajectory. Our massless charges are actually moving slower than light. But since $\gamma$ is given by the ratio of macroscopic and microscopic quantities, in all possible applications of EMC one has $\gamma \gg 1$, allowing to approximate the actual velocity by (\ref{vel}).}.

The full EMC system of equations is, in the 3+1 split, 
\begin{equation}\label{max1}
\dot{{\bf B}}=-\nabla \times {\bf E},
\end{equation}
\begin{equation}\label{max2}
\dot{{\bf E}}=\nabla \times {\bf B}-\rho _+{\bf v}_++\rho _-{\bf v}_-,
\end{equation}
\begin{equation}\label{con}
\dot{\rho_\pm }+\nabla\cdot(\rho_\pm{\bf v}_\pm )=Q.
\end{equation}
Here $\rho_\pm$ are the particle densities (multiplied by the absolute value of the charge), and the source $Q$ represents pair creation/annihilation. The source $Q$ can be chosen at will. This is because our massless particles do not carry energy and momentum (again because of the strong radiation damping), and therefore pairs can be created and annihilated arbitrarily. The only requirement is that $Q\geq 0$ if one of the densities $\rho_\pm$ vanishes -- there can be no negative particle densities. 

Somewhat surprisingly, even in this incomplete form, EMC seems to be useful, because for a broad range of assumed sources $Q$, the EMC predictions remain virtually unchanged, and one can indeed choose $Q$ as he wishes. In pulsar applications, an arbitrarily chosen $Q$ allows to calculate only the bolometric luminosity, while for the spectra and light curves one must calculate the source $Q$ from the actual charge-photon kinetics. This means that the photon distribution function needs to be added to the EMC system. The important point, however, is that even in this full pulsar theory, the field and charge dynamics is still given by the EMC system (\ref{max1}-\ref{con}). 

Applicability of EMC to pulsars requires instantaneous parallel dynamics at light cylinder, which gives
\begin{equation}\label{app}
L\gg L_e\left({R\over r_e}\right)^{2/3},
\end{equation}
where $L$ is the spin-down luminosity, $R$ is the light cylinder radius, $r_e={e^2\over mc^2}$ is the classical electron radius, $L_e={mc^3\over r_e}= 8.7\times 10^{16}$erg/s is the "classical electron luminosity". Equation (\ref{app}), for a 30ms pulsar, gives $L\gg 6\times 10^{30}$erg/s, and is easily satisfied. 

\section{Aligned rotator in EMC}

To illustrate how EMC works in pulsar applications (the only application the author is aware of), we simply assume that $Q$, whatever it is, creates a prefixed quantity of high-multiplicity plasma, meaning
\begin{equation}\label{ans}
\rho_++\rho_-=|\rho| +f\rho_0,
\end{equation}
Where $\rho =\rho_+-\rho_-$ is the charge (Goldreich-Julian) density,  $f$ is the multiplicity and $\rho_0\gtrsim |\rho |$ is a fiducial charge density greater than the absolute value of the actual charge density $|\rho|$.

\begin{figure}[h]
   \centering
   \includegraphics[width=3.3in]{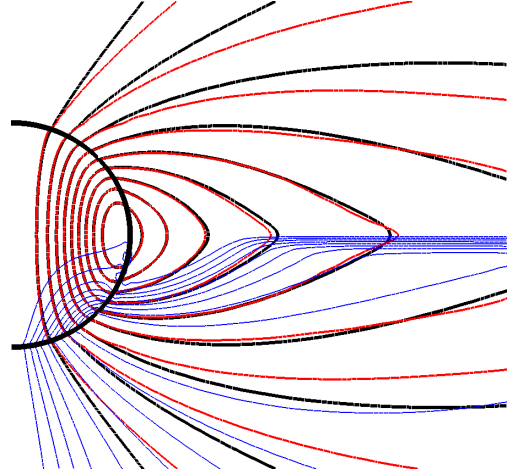} 
   \caption{Isolines of (i) the poloidal magnetic flux -- thick black, (ii) the electric potential -- thinner red, (iii) the poloidal current  -- thin blue (shown only below the equator for clarity of the picture). }
   \label{fig1}
\end{figure}

In the figure we show the results for $f=10$ and $\rho_0 =\sqrt{B^2+E^2}/R$, where $R$ is the spherical radius. This $\rho_0$ simply represents the charge density which is capable of changing the fields by an order unity factor. The simulation technique is similar to what we have used when simulating pulsars by Strong-Field Electrodynamics (SFE)  (Gruzinov 2011a). The light cylinder is at twice the stellar radius. 

We get an absolute replica of the pulsar magnetosphere calculated in the high-conductivity  limit of SFE. This calls for an explanation.

\section{EMC and SFE}

SFE is Maxwell theory plus Lorentz-covariant Ohm's law (Gruzinov 2011a and references therein)
\begin{equation}\label{sfe}
{\bf j}={\rho {\bf E}\times {\bf B}+\sqrt{\rho ^2+\gamma ^2\sigma ^2 E_0^2}(B_0{\bf B}+E_0{\bf E})\over B^2+E_0^2},
\end{equation}
\begin{equation}
 \gamma ^2\equiv {B^2+E_0^2\over B_0^2+E_0^2}.
\end{equation}
Here $\sigma$ is the conductivity, which in SFE is assumed to be an arbitrary scalar depending on $E_0$ and $B_0$. This Ohm's law has been postulated just from the Lorentz covariance requirement. 

SFE can be derived from EMC if one postulates the Lorentz-invariant source term
\begin{equation}\label{Q}
Q=\kappa(\sigma ^2E_0^2-2j^\mu_+j_{\mu -}),
\end{equation}
where $j^\mu_\pm $ are the Lorentz-covariant 4-currents $j^\mu_\pm =\rho_\pm(1,{\bf v}_\pm)$. In the limit $\kappa \rightarrow \infty$, in the frame where the fields are parallel and the charge density vanishes, the particle densities become $\rho_\pm=\sigma E_0/2$ and we do get the current (\ref{sfe}).

It is good to know that SFE (a simpler but postulated theory) is derivable from EMC (a first-principle theory, but with more fields and not fully defined).  This does not explain, however,  why an arbitrarily chosen high-multiplicity plasma of \S2 gives the same magnetosphere as the high-conductivity SFE.

The universality of the high-conductivity SFE magnetosphere follows from the expression for the electric current in EMC:
\begin{equation}\label{nmcc}
{\bf j}={\rho {\bf E}\times {\bf B}+(\rho_++\rho_-)(B_0{\bf B}+E_0{\bf E})\over B^2+E_0^2}.
\end{equation}
Now comparing the EMC current (\ref{nmcc}) to the SFE current (\ref{sfe}), we see that the equilibrium state of the high-multiplicity plasma of \S2 is identical to the SFE equilibrium with a certain (weird) space-dependent conductivity -- but as long as the conductivity is everywhere high, its actual value is irrelevant.

The following seems to be a fair analogy. The temperature jump at the ideal gas shock can be calculated in dissipative gas dynamics with arbitrary but small viscosity. Similarly, the amount of damping of an ideal pulsar magnetosphere can be  calculated in EMC with arbitrary but high multiplicity, or in SFE  with arbitrary but high conductivity. 

EMC and SFE give identical results for the bolometric luminosity. But only EMC seems to be a portion of the full pulsar theory, because EMC gives a faithful description of the particle density (except for creation and annihilation if any). 

\section{Summary}

EMC -- Electrodynamics of Massless Charges -- is a well-behaved dynamical system. The charges carry zero energy-momentum and obey Aristotelean dynamics. The system is semi-dissipative. The electromagnetic energy can decrease, but for certain field configurations it is precisely conserved. 

For pulsars, in the high-multiplicity limit, EMC gives an undamped Poynting flux everywhere except at the equatorial current layer. As the charges carry no energy, the equatorial Poynting flux damping must be interpreted as  bolometric luminosity. For an aligned rotator, one gets the bolometric luminosity $\approx  50\%$ of the spin-down power (Gruzinov 2011b). For a generic inclination angle, the bolometric luminosity has not yet been calculated. Presumably the bolometric luminosity at generic inclination is smaller than 50\% of the spin-down power, because the singular current layer is weaker in the inclined case (Spitkovsky 2006).

\acknowledgements

I thank Matt Kleban for useful discussions.


\begin{references}

\reference{} Gruzinov, A., 2011a, Pulsar Magnetosphere, arXiv:1101.3100

\reference{} Gruzinov, A., 2011b, Ohmic Power of Ideal Pulsars, arXiv:1101.5844 

\reference{} Spitkovsky, A., 2006, Time-dependent Force-free Pulsar Magnetospheres: Axisymmetric and Oblique Rotators, ApJ, 648, L51



\end{references}
\end{document}